# Clustering global ocean profiles according to temperature-salinity structure

**Nozomi Sugiura**[1,*]

[1]Research and Development Center for Global Change, Japan Agency for Marine-Earth Science and Technology, Yokosuka, Japan
[*]nsugiura@jamstec.go.jp

## ABSTRACT

An unsupervised clustering using a Gaussian mixture model is applied to Argo profiles distributed over the entire ocean. We employ as the coordinate components in feature space the path signature, which is a central notion in rough path theory. This allows us to better identify the oceanic condition at each horizontal point with distinct clusters, than by using conventional temperature and salinity coordinate. To the best of my knowledge, it is the first attempt at clustering almost all of the existing Argo profiles with the full use of measured sequences of temperature, salinity, and pressure. We will also discuss why the path signature is relevant to representing the property of a profile.

## Introduction

Argo is an international effort involving the collecting of high-quality temperature and salinity profiles, typically from the upper 2000 m of the global ocean[1]. To enhance the understanding of ocean and climate dynamics, it is of importance to classify the profiles into clusters that represent distinct properties. Although it would be desirable for clustering to cover a larger number of samples over a wider area of ocean, the complicated shape of profiles have prevented us from an unified procedure to complete a global mapping.

As an early work, Hjelmervik and Hjelmervik (2013)[2] used PCA for the slices of vertical Temperature and Salinity sequence for ocean front detection. Then, the first successful attempt to classify the ocean with the Gaussian mixture model (GMM) has been performed by Maze et al. (2017)[3]. They found a geometrically coherent pattern of clusters in their target area, i.e. the Atlantic ocean. They also contributed to provide a toolbox for the GMM clustering[4] in public. Afterwards, Houghton and Wilson (2019)[5] classified temperature profiles for the use of El nino Prediction. Jones et al. (2019)[6] applied GMM to temperature profiles to study the Southern ocean. Boehme and Rosso (2021)[7] used PCA for the slices of vertical Temperature and Salinity sequence, and applied GMM for a regional analysis.

In conventional approaches, each levels of the vertical axis of each ocean variables (Temperature, Salinity, or both of them), or total heat content in the water column only, are considered a feature (e.g., Maze et al. (2017a)[4]).

Our approach is more comprehensive than previous studies in the sense that all the profiles of Temperature and Salinity in the global ocean are subject to the GMM classification. One key point for enabling this is that we use the signature of each profile as feature set, instead of the vertical slices of a profile.

Consider a curve that is represented by a sequence of vector consisting Pressure, Salinity, and Temperature: $X = \{X_t \in \mathbb{R}^d | 0 \leq t \leq 1, d = 3\}$. We can naively slice and sample points in the curve: $\tilde{X} = \{X_{t_i} \in \mathbb{R}^d | i = 1, 2, \cdots, T\}$. Then, a continuous function of this curve can be approximately represented as

$$f(X) \approx \tilde{f}(\tilde{X}), \tag{1}$$

with a function $\tilde{f}$ of $\tilde{X}$. However, a problem here is that such function $\tilde{f}$ is in general complicated and nonlinear. This nonlinearity issue cannot be resolved even if we apply $X$ a Fourier transform, which is just a linear operation. On the other hand, we can also describe the curve with its signature $\mathcal{S}_n(X)$ (see Methods for definition). Then, by universal approximation theorem[8], any nonlinear function $f$ of the curve $X$ can be approximately represented as

$$f(X) \approx \left\langle \overline{f}, \mathcal{S}_n(X) \right\rangle, \tag{2}$$

where $\overline{f}$ is a linear functional. By comparing expressions (1) and (2), we see that transforming a curve into its signature has an advantage in expressing any function of curves linearly. In Eq. (2), nonlinearity has been absorbed in the signature. This stems

from the property that the product of two linear functionals is another linear functional:

$$(f \cdot g)(X) \equiv \langle \overline{f}, \mathcal{S}_n(X) \rangle \langle \overline{g}, \mathcal{S}_n(X) \rangle = \langle \overline{f} \shuffle \overline{g}, \mathcal{S}_n(X) \rangle, \tag{3}$$

where $\shuffle$ represents the shuffle product[9].

By exploiting the signature as a feature of each profile, we perform a clustering of Argo profiles in the global ocean, using observation of both Temperature and Salinity.

## Methods

### Extraction of Feature

We use the snapshot Argo data as of December 2020[10], which comprises about $1 \times 10^6$ profiles. A profile observation is described as a sequence of $d = 3$-dimensional vectors consisting of Pressure, Salinity, and Temperature: $P_t = X_t^1, S_t = X_t^2, T_t = X_t^2$. Each profile $\{X_t \in \mathbb{R}^d | 0 \leq t \leq 1\}$ is transformed into the path signature, which is the set of all the iterated integrals:

$$\mathsf{X}^{(i_1 \cdots i_n)} = \mathcal{I}_n(X)^{(i_1 \cdots i_n)} = \int_{t_n=0}^{1} \cdots \int_{t_1=0}^{t_2} dX_{t_1}^{i_1} \cdots dX_{t_n}^{i_n} \in \mathbb{R}, \tag{4}$$

$$\mathcal{I}_j(X) = \{\mathsf{X}^{(i_1 \cdots i_j)}\}_{(i_1 \cdots i_j) \in \{1, \cdots, d\}^j} \in (\mathbb{R}^d)^{\otimes j}, \tag{5}$$

$$\mathcal{S}_n(X) = (1, \mathcal{I}_1(X), \cdots, \mathcal{I}_n(X)) \in (\mathbb{R}^d)^{\otimes 0} \oplus (\mathbb{R}^d)^{\otimes 1} \oplus \cdots \oplus (\mathbb{R}^d)^{\otimes n}. \tag{6}$$

The truncated signature up to $n$ degree, $\mathcal{S}_n(X)$, has $D_0 = (d^{n+1} - 1)/(d-1)$ components in total. For detail, refer to Appendix A or Sugiura and Hosoda (2020)[11].

Signature only contains information relative to the starting point. To take into account the absolute value of Salinity and Temperature, the feature $x$ of each profile is augmented with their absolute values of Salinity $X_e^2$ and Temperature $X_e^3$ at endpoint Pressure $X_e^1 = 2000$.

$$x = \{X_e^2, X_e^3\} \oplus \mathcal{S}_n(X) \in \mathbb{R}^D, \tag{7}$$

where $D = D_0 + 2$.

### Gaussian Mixture Model

Suppose we have data represented by feature set $\{x_i \in \mathbb{R}^D\}_{i=1}^N$. Probability density at value $x_i \in \mathbb{R}^D$ for a Gaussian mixture model (GMM) with $M$ clusters is defined as

$$p(x_i|\theta) = \sum_{m=1}^{M} \lambda_m \mathcal{N}(x_i|\mu_m, \Sigma_m), \tag{8}$$

$$\mathcal{N}(x_i|\mu_m, \Sigma_m) = |2\pi\Sigma_m|^{-\frac{1}{2}} \exp\left(-\frac{1}{2}(x_i - \mu_m)^T \Sigma_m^{-1}(x_i - \mu_m)\right), \tag{9}$$

where $|\cdot|$ denotes the determinant of matrix, and $\theta = \{\lambda_m, \mu_m, \Sigma_m\}_{m=1}^M$ is the set of parameters for the model. Note that the assignment of cluster to each profile, $z_i$, is treated as a latent random variable, which can not be observed.

Then, the log-likelihood and Bayesian Information Criterion (BIC) of the parameter $\theta$ given data $\{x_i\}_{i=1}^N$ are defined as

$$\ell(\theta) = \sum_{i=1}^{N} \log p(x_i|\theta), \tag{10}$$

$$\text{BIC}(\theta) = -2\ell(\theta) + K \log N, \tag{11}$$

where $N$ is the number of data, $K$ is the degrees of freedom in $\theta$, that is $K = M - 1 + M(D + D(D+1)/2)$ because mean $\mu_m$ is in $\mathbb{R}^D$, covariance $\Sigma_m \in \mathbb{R}^{D \times D}$ is a symmetric positive semi-definite matrix, and weights $\lambda_{1:M}$ satisfy $\sum_{m=1}^{M} \lambda_m = 1$.

In clustering, an expectation-maximization algorithm is used to find a set of parameters that maximize the log-likelihood $\ell$. This allows us to determine to which cluster each profile is most likely to belong. Furthermore, the GMM also serves as a generative model for the profiles. Note that an appropriate number of cluster, $M$. is determined by comparing several settings and finding the setting with minimum BIC value. For detail, refer to Maze et al. (2017b)[3].

## Results

Figure 1 compares the BIC values for the results with different number of clusters. Although the values for 6 to 9 clusters are comparable, we choose 7, which has the minimum BIC. Figure 2 shows the clusters in the signature space projected onto 2-dimensional planes. The axes are the commutators for the second iterated integrals, e.g., $[P, T] = \int_0^1 \int_0^{u_2} (dP_{u_1} dT_{u_2} - dT_{u_1} dP_{u_2})$. It it clear that each cluster is coherent in the signature space. Although these marginal distributions should be Gaussian, they do not perfectly look like Gaussian.

In fig 3 we show the global distribution of profiles each of which is colored with the most probable clusters it belongs to. Figures 5 to 11 show the geographical distributions of samples in clusters. The profiles in the map measured in boreal summer are indicated with bright colors, otherwise dark colors. TS-profiles for the centroid and samples from profile data are also shown in the right figures.

Figure 12 shows samples generated by the Gaussian Mixture Model on $T$-$S$ diagram. Each sample of signature is generated according to the probability assigned by the GMM. Then, by minimizing a cost function of the form (23), the corresponding profile is derived.

## Discussion

We have successfully classified almost all of the historical Argo profiles retrieved from the entire ocean, with the use of both Temperature and Salinity measurements.

By comparing the BIC values for different numbers of clusters, we chose 7 clusters as a candidate of representing the ocean.

These clusters seem to well represent major oceanic and climatic conditions including subtropical circulation, subarctic circulation, subtropical mode waters, weak stratification, high precipitation, strong evaporation, and river discharge.

In this study, we used a Maharanobis distance of signatures, which corresponds to Euclidean kernel in the signature space. Since each component signature is not Gaussian (see Fig. 2), the GMM setting may not be perfectly effective. If we change the kernel to the one that takes into account nonlinearity in the signature space, the Gaussian modeling could work better. For that purpose, kernel methods[12] will be useful, provided that we overcome the computational issue regarding the huge number of samples in the data.

## References

1. Gould, J. *et al.* Argo profiling floats bring new era of in situ ocean observations. *Eos, Transactions Am. Geophys. Union* **85**, 185–191, DOI: 10.1029/2004EO190002 (2004). https://agupubs.onlinelibrary.wiley.com/doi/pdf/10.1029/2004EO190002.

2. Hjelmervik, K. T. & Hjelmervik, K. Estimating temperature and salinity profiles using empirical orthogonal functions and clustering on historical measurements. *Ocean. Dyn.* **63**, 809–821, DOI: 10.1007/s10236-013-0623-3 (2013).

3. Maze, G. *et al.* Coherent heat patterns revealed by unsupervised classification of argo temperature profiles in the north atlantic ocean. *Prog. Oceanogr.* **151**, 275 – 292, DOI: https://doi.org/10.1016/j.pocean.2016.12.008 (2017).

4. Maze Guillaume, C. C., Mercier Herle. Profile classification models (2017).

5. Houghton, I. A. & Wilson, J. D. El Niño Detection Via Unsupervised Clustering of Argo Temperature Profiles. *J. Geophys. Res. Ocean.* **125**, e2019JC015947, DOI: https://doi.org/10.1029/2019JC015947 (2020). E2019JC015947 10.1029/2019JC015947, https://agupubs.onlinelibrary.wiley.com/doi/pdf/10.1029/2019JC015947.

6. Jones, D. C., Holt, H. J., Meijers, A. J. S. & Shuckburgh, E. Unsupervised Clustering of Southern Ocean Argo Float Temperature Profiles. *J. Geophys. Res. Ocean.* **124**, 390–402, DOI: https://doi.org/10.1029/2018JC014629 (2019). https://agupubs.onlinelibrary.wiley.com/doi/pdf/10.1029/2018JC014629.

7. Boehme, L. & Rosso, I. Classifying Oceanographic Structures in the Amundsen Sea, Antarctica. *Geophys. Res. Lett.* **48**, e2020GL089412, DOI: https://doi.org/10.1029/2020GL089412 (2021). E2020GL089412 2020GL089412, https://agupubs.onlinelibrary.wiley.com/doi/pdf/10.1029/2020GL089412.

8. Fermanian, A. Embedding and learning with signatures. *Comput. Stat. Data Analysis* **157**, 107148, DOI: https://doi.org/10.1016/j.csda.2020.107148 (2021).

9. Chevyrev, I. & Kormilitzin, A. A Primer on the Signature Method in Machine Learning. *ArXiv e-prints* (2016). 1603.03788.

10. Argo. Argo float data and metadata from Global Data Assembly Centre (Argo GDAC) - Snapshot of Argo GDAC of December 10th 2020, DOI: https://doi.org/10.17882/42182#79118 (2020).

11. Sugiura, N. & Hosoda, S. Machine learning technique using the signature method for automated quality control of argo profiles. *Earth Space Sci.* **7**, e2019EA001019, DOI: https://doi.org/10.1029/2019EA001019 (2020). E2019EA001019 10.1029/2019EA001019, https://agupubs.onlinelibrary.wiley.com/doi/pdf/10.1029/2019EA001019.

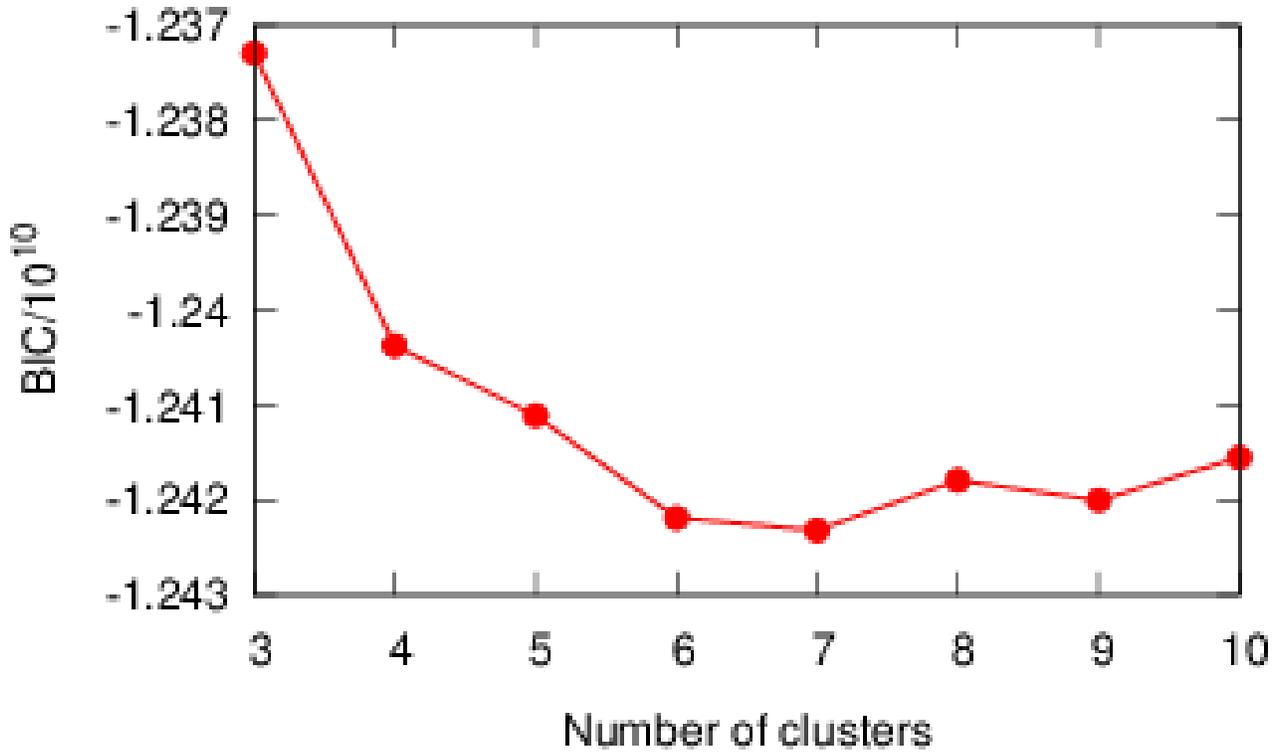

**Figure 1.** BIC values according to the number of groups.

12. Kiraly, F. J. & Oberhauser, H. Kernels for sequentially ordered data. *J. Mach. Learn. Res.* **20**, 1–45 (2019).

## Acknowledgements
This study is funded by JST-PROJECT-20218919.

| Cluster number | Properties |
|---|---|
| Cluster 1 | Areas involved in subtropical circulation |
| Cluster 2 | Areas involved in Mediterranean water or subtropical mode waters |
| Cluster 3 | High evaporation areas |
| Cluster 4 | Areas involved in subarctic circulation or Mediterranean Sea |
| Cluster 5 | Areas where freshwater from the estuary is distributed in the surface layer |
| Cluster 6 | High precipitation zone |
| Cluster 7 | Areas with weak stratification: transition regions or Antarctic circumpolar current |

**Table 1.** Properties of the clusters.

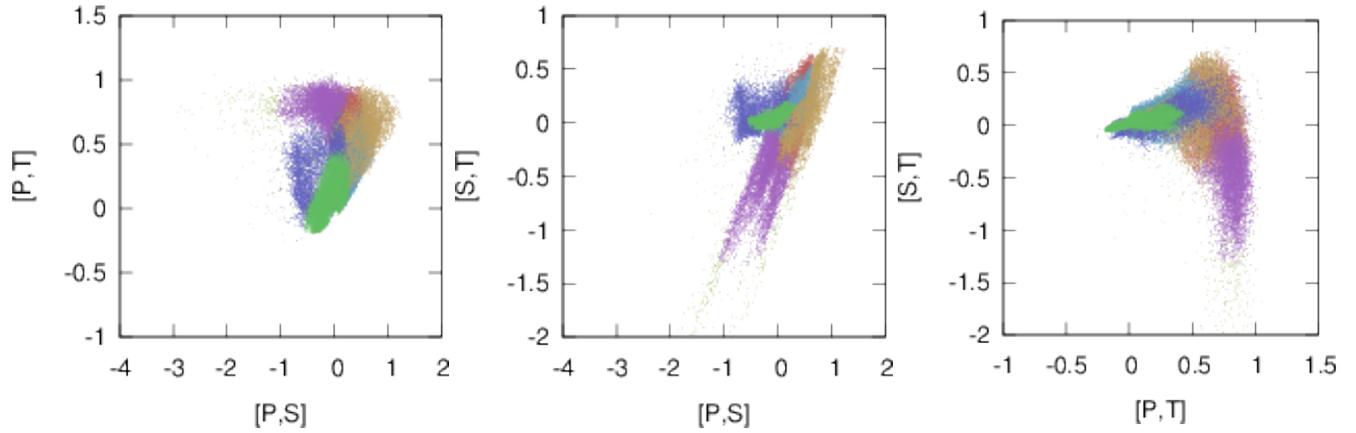

**Figure 2.** Distributions of clusters projected onto $[P,S]$–$[P,T]$, $[P,S]$–$[S,T]$, and $[P,T]$–$[S,T]$ plane. Clusters are numbered and colored as **1**, **2**, **3**, **4**, **5**, **6**, and **7**.

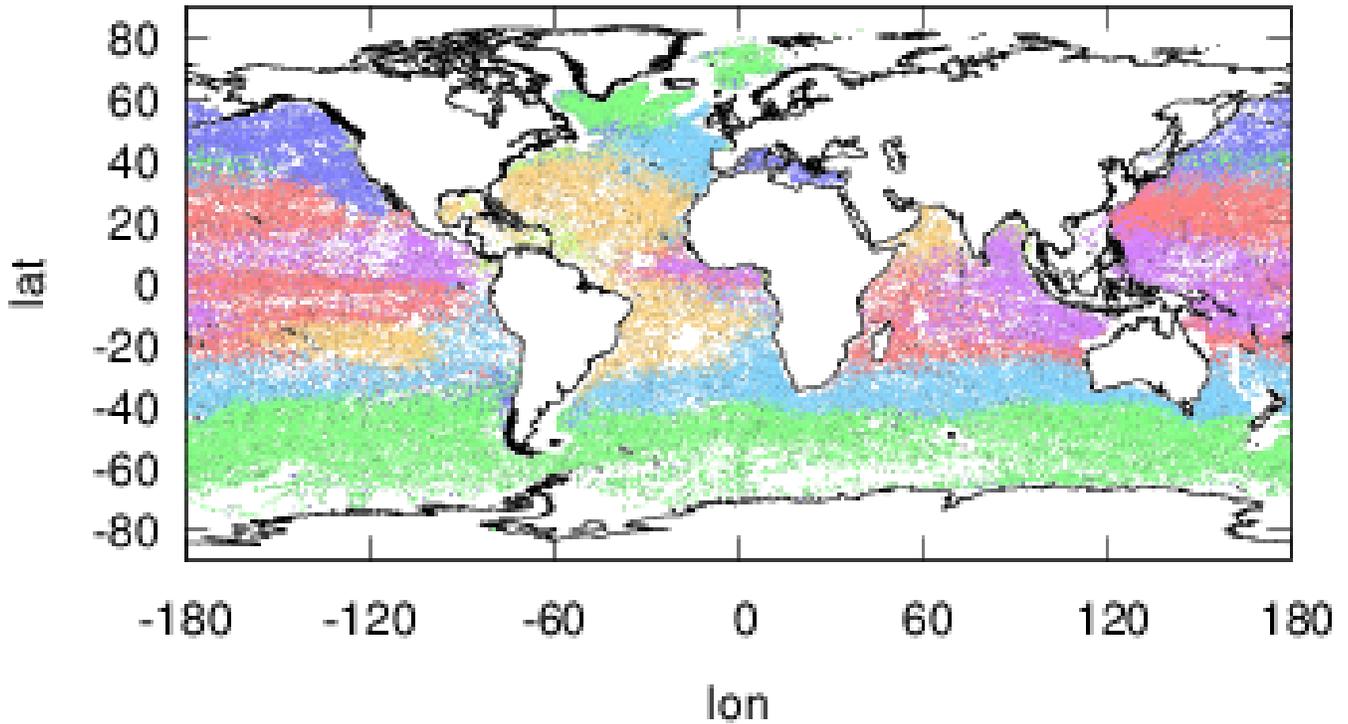

**Figure 3.** Global distribution of profiles each of which is colored with the most probable clusters it belongs to. Clusters are colored as **1**, **2**, **3**, **4**, **5**, **6**, and **7**.

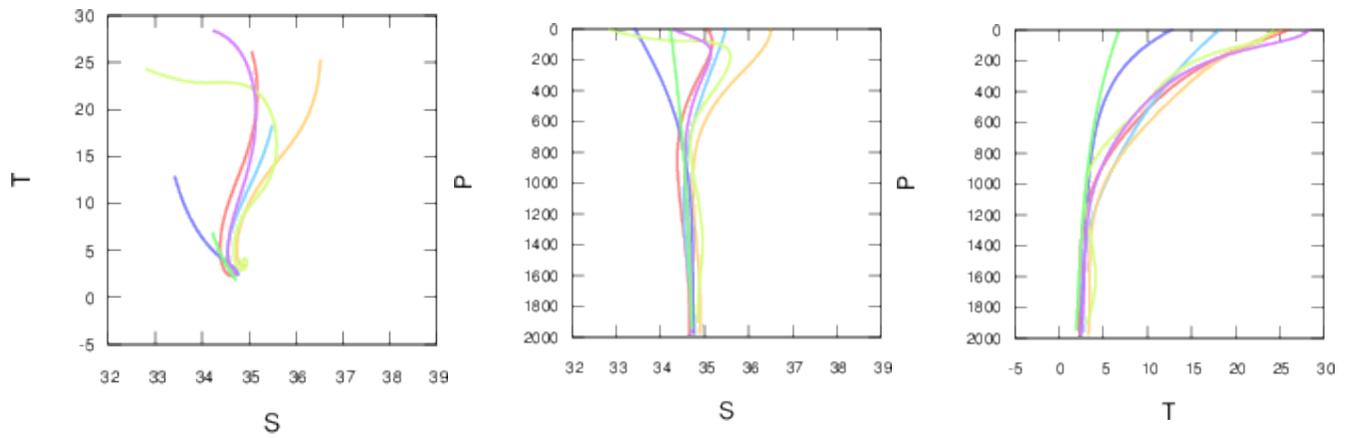

**Figure 4.** Profile corresponding to the centroid for each cluster. Colored as 1, 2, 3, 4, 5, 6, and 7.

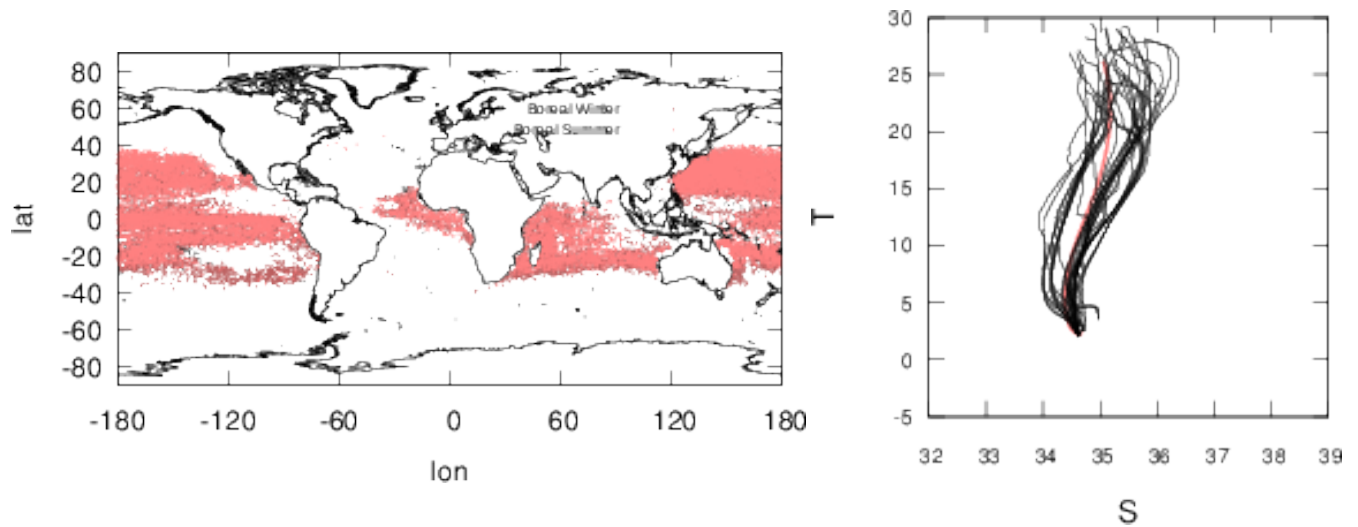

**Figure 5.** Left: Geographical distribution of Cluster 1 (areas involved in subtropical circulation). The profiles in the map measured in boreal summer (May to October) are indicated with bright colors, otherwise dark colors. Right: TS-profiles for the centroid (color) and samples from profile data (black).

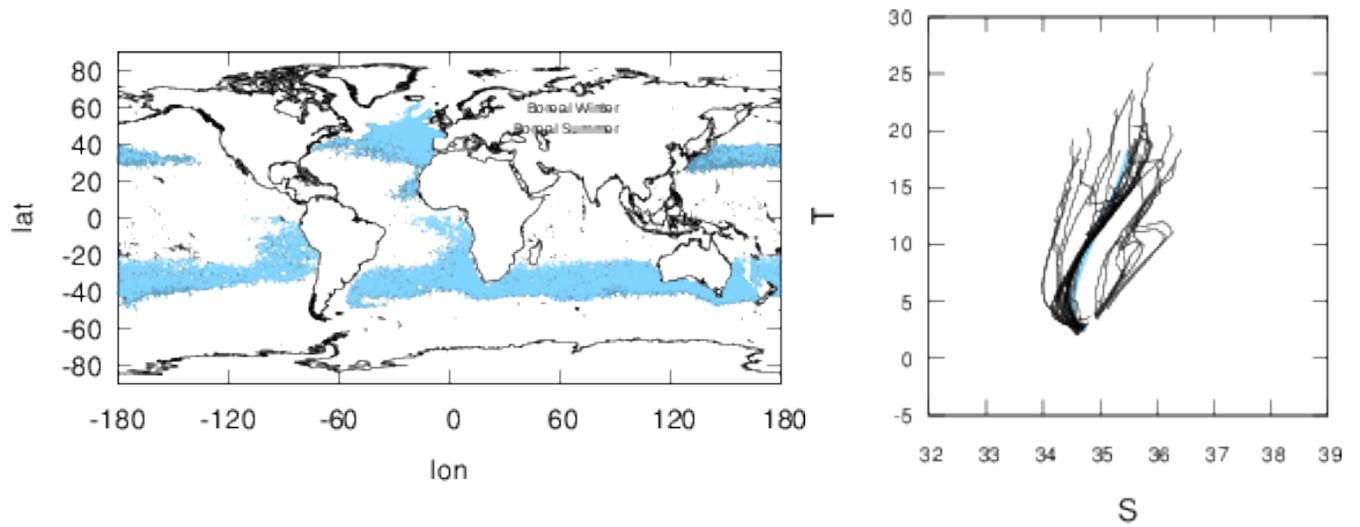

**Figure 6.** Left: Geographical distribution of Cluster 2 (areas involved in Mediterranean water or subtropical mode waters). The profiles in the map measured in boreal summer are indicated with bright colors, otherwise dark colors. Right: TS-profiles for the centroid (color) and samples from profile data (black).

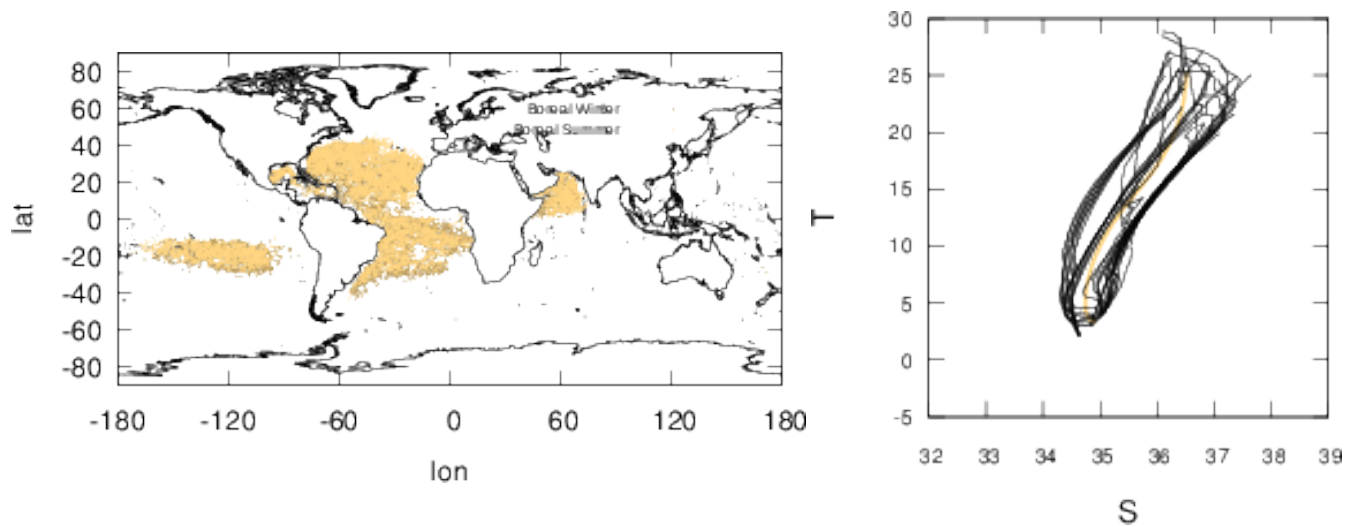

**Figure 7.** Left: Geographical distribution of Cluster 3 (high evaporation areas). The profiles in the map measured in boreal summer are indicated with bright colors, otherwise dark colors. Right: TS-profiles for the centroid (color) and samples from profile data (black).

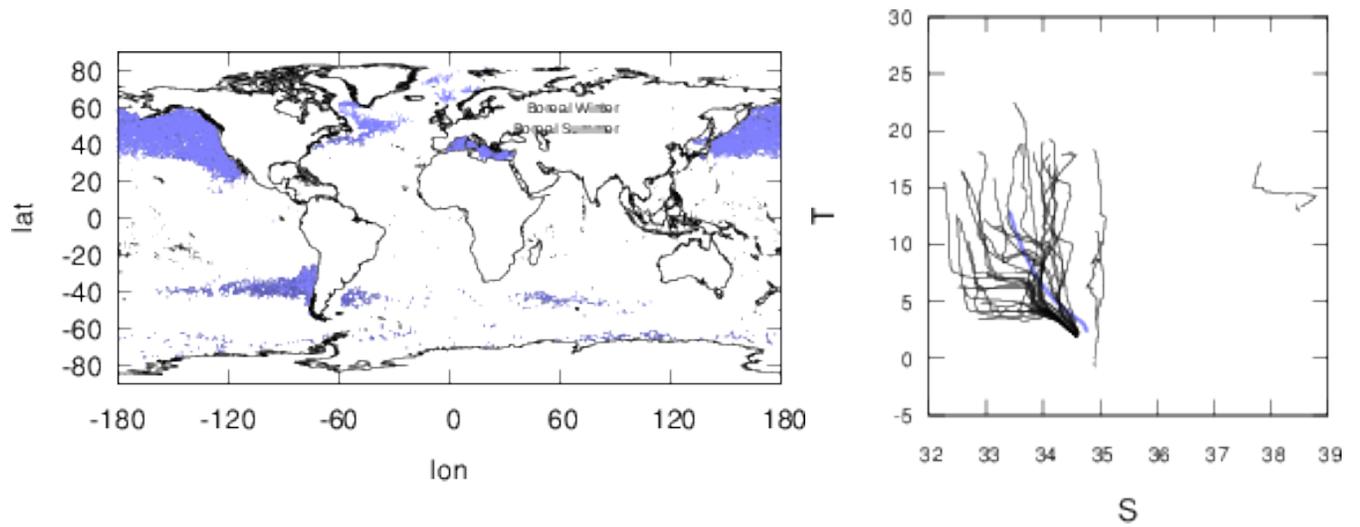

**Figure 8.** Left: Geographical distribution of Cluster 4 (areas involved in subarctic circulation or Mediterranean Sea). The profiles in the map measured in boreal summer are indicated with bright colors, otherwise dark colors. Right: TS-profiles for the centroid (color) and samples from profile data (black).

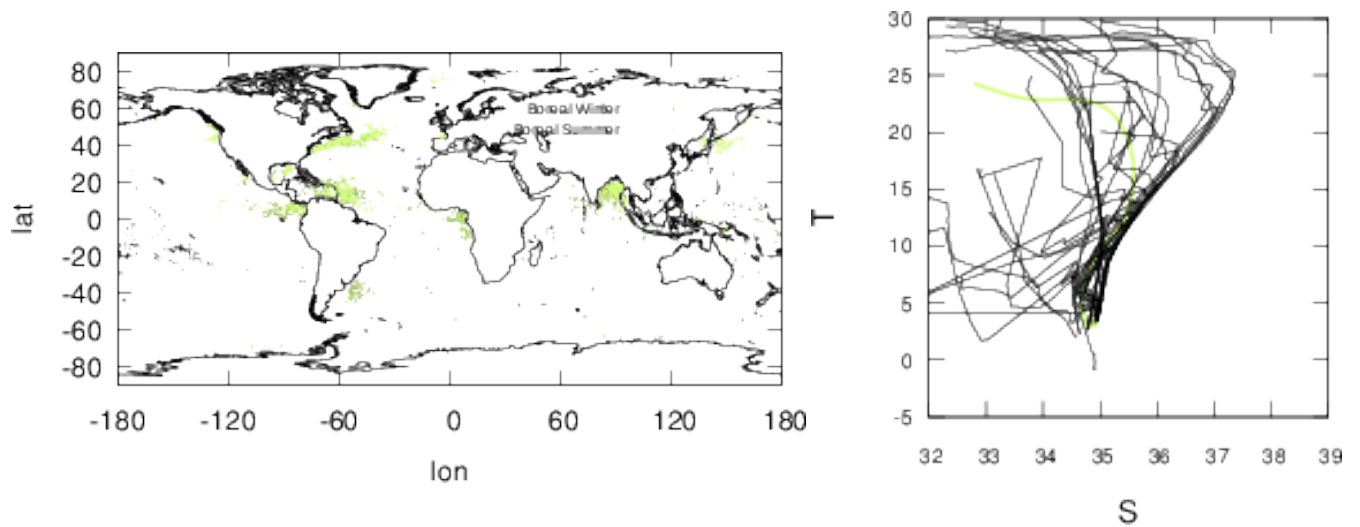

**Figure 9.** Left: Geographical distribution of Cluster 5 (areas where freshwater from the estuary is distributed in the surface layer). The profiles in the map measured in boreal summer are indicated with bright colors, otherwise dark colors. Right: TS-profiles for the centroid (color) and samples from profile data (black).

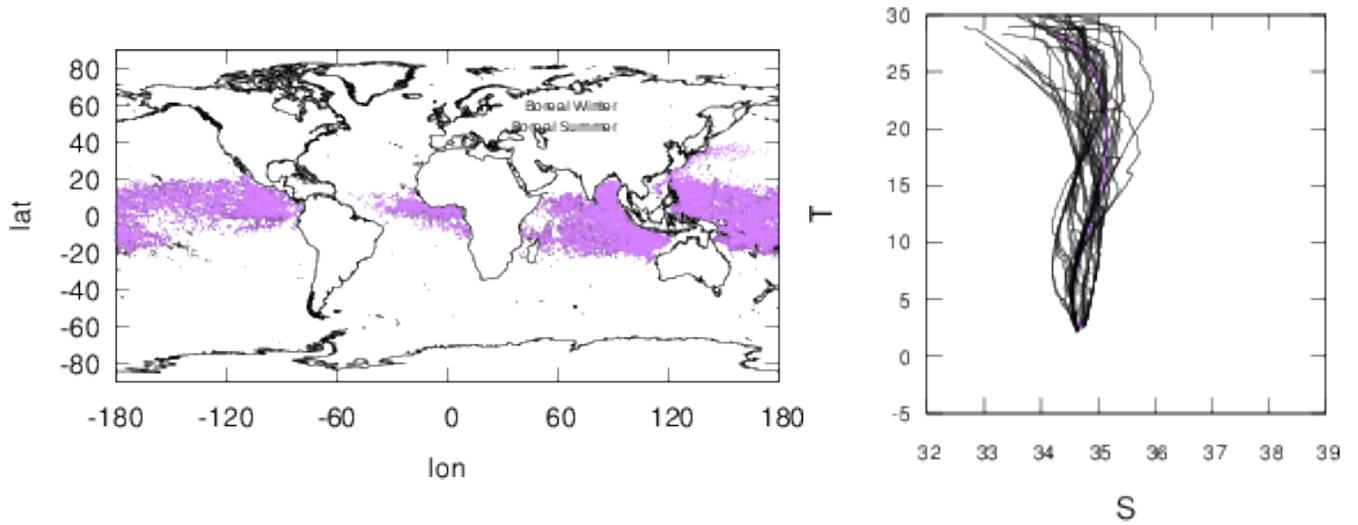

**Figure 10.** Left: Geographical distribution of Cluster 6 (high precipitation zone). The profiles in the map measured in boreal summer are indicated with bright colors, otherwise dark colors. Right: TS-profiles for the centroid (color) and samples from profile data (black).

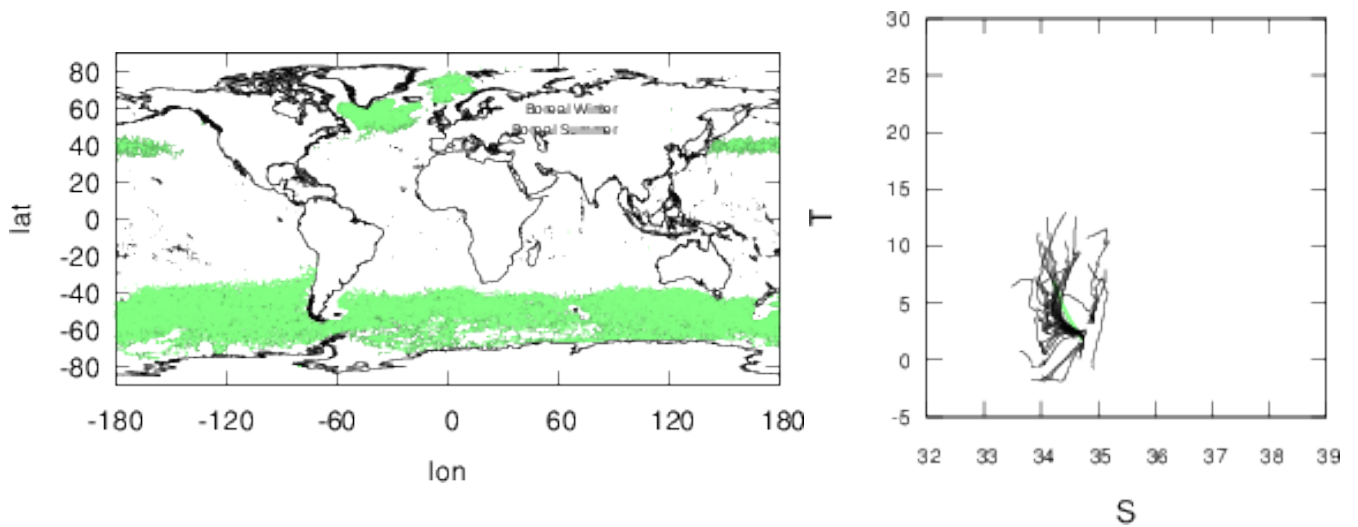

**Figure 11.** Left: Geographical distribution of Cluster 7 (areas with weak stratification: transition regions or Antarctic circumpolar current). The profiles in the map measured in boreal summer are indicated with bright colors, otherwise dark colors. Right: TS-profiles for the centroid (color) and samples from profile data (black).

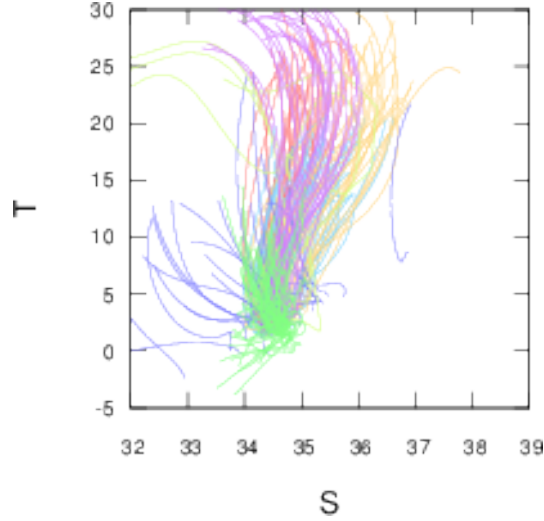

**Figure 12.** Samples generated by the Gaussian Mixture Model. Each profile is colored with the cluster it belongs to as 1, 2, 3, 4, 5, 6, and 7.

## A Rationale for using signature

We consider the situation where each point in a path belongs to underlying space $V = \mathbb{R}^d$. Truncated tensor algebra $T_n(V) = \bigoplus_{0 \le k \le n} V^{\otimes k}$ is the sum of all tensor product of $V$ up to order $n$, equipped with multiplication: $a \otimes b = (c_0, c_1, \cdots)$, $c_k = \sum_{i=0}^{k} a_i \otimes b_{k-i}$, as well as addition, and scalar multiplication. We write the basis for $V^{\otimes p}$ as $e_{i_1 \cdots i_p} = e_{i_1} \otimes \cdots \otimes e_{i_p}$. Signature is an element of $T_n(V)$ that has a form of

$$\mathcal{S}_n(X) := 1 + \mathcal{I}_1(X) + \cdots + \mathcal{I}_n(X), \tag{12}$$

$$\mathcal{I}_r(X) := \sum_{|w|=r} \int dX^{(w)} e_w, \tag{13}$$

$$\int dX^{(w)} := \int_{0 < u_1 < \cdots < u_r < 1} dX^{i_1}_{u_1} \cdots dX^{i_r}_{u_r}, \tag{14}$$

where the multi-index $w = (i_1, \cdots, i_r)$ with length $|w| = r$ runs across $\{1, \cdots, d\}^r$. We introduce an inner product on $T_n(V)$ that obeys

$$(e_I, e_J) = (\Sigma^{-1})_{IJ} \tag{15}$$

where $I$ and $J$ are multi-indices, and $\Sigma \in \mathbb{R}^{N \times N}, N = 1 + d + \cdots + d^n$ is a covariance matrix. Then for paths $X$ and $Y$, we have

$$k(X,Y) := (\mathcal{S}_n(X), \mathcal{S}_n(Y)) = \sum_{r=0}^{n} \sum_{w_r, w'_r} \int dX^{(w_r)} (\Sigma^{-1})_{w_r w'_r} \int dY^{(w'_r)} \tag{16}$$

$k(X,Y)$ is a positive definite kernel because $\sum_{ij} \alpha_i \alpha_r k(X_i, X_r) = (\sum_i \alpha_i \mathcal{S}(X_i), \sum_r \alpha_r \mathcal{S}(X_r)) \ge 0$. Thus, we can regard that the clustering is performed on a reproducing kernel Hilbert space $\mathcal{H}$ that consists of functions $f$ on the space of curves:

$$f(\cdot) = \sum_{i=1}^{N} f_i k(\cdot, X_i). \tag{17}$$

An inner product of functional, $f$ and $g$, is naturally defined as

$$(f,g) = \sum_{ij} f_i g_j k(X_i, X_j) = \sum_{ij} f_i g_j \big(\mathcal{S}_n(X_i), \mathcal{S}_n(X_j)\big) \tag{18}$$

Also, the distance between functions $f, g \in \mathcal{H}$ is defined as $d(f,g)^2 = (f-g, f-g)$. By Cauchy–Schwartz inequality, we have

$$|f(X) - f(Y)| \le d(f,0) d(X,Y), \tag{19}$$

On the other hand, a continuous function $F$ of paths is arbitrary well approximated by a linear functional $f$ (e.g., Proposition 3 of Fermaninan (2020)[8]):

$$\forall X) \quad |F(X) - f(X)| < \epsilon \tag{20}$$

Estimates (19) and (20) imply a uniform estimate: for all $\epsilon > 0$, there exits $\delta > 0$ such that

$$\forall X, Y \quad d(X,Y) < \delta \implies |F(X) - F(Y)| < \epsilon. \tag{21}$$

If we interpret the function $F$ as an assignment of "property" to a path, this means that the properties of any two arbitrary chosen paths are close in a neighborhood in the signature space. It thus makes sense to perform clustering of the profiles on space $\mathcal{H}$, where we take the signature of each profile as feature set.

## B Practical reconstruction of a path from a graded tensor

Suppose we have the mean, or centroid, of signatures in a cluster, which is expressed as a linear combination of the truncate signatures up to order $n$ the data:

$$\mu = \sum_{i=1}^{N} \gamma_i x_i \in \mathbb{R}^D, \tag{22}$$

where $x_i$ is the signature for the $i$-th profile, and $\mu \in \mathbb{R}^D$ is a mean of signatures, and $\gamma_i \ge 0$ is the coefficient. We have to be careful that $\mu$, which can be seen as a function on path space (see appendix A), does not necessarily represent a path. In other words, iterated integrals in $\mu$ might not strictly satisfy the shuffle identity: for example $\mu^{(1)} \mu^{(2)} = \mu^{(12)} + \mu^{(21)}$. Thus, the proper problem setup here is to find a path $X$ that has a signature close to $\mu$. Assuming that the path $X$ is a polyline connecting points $X_1, \cdots, X_m \in \mathbb{R}^d$, cost function $J$ is defined as

$$J(X) = \frac{1}{2} \sum_{i_1=1,\cdots,d} \left( \mathcal{I}_1(X)^{(i_1)} - \mu^{(i_1)} \right)^2 + \cdots + \frac{1}{2} \sum_{i_1,\cdots,i_n=1,\cdots,d} \left( \mathcal{I}_n(X)^{(i_1,\cdots,i_n)} - \mu^{(i_1,\cdots,i_n)} \right)^2 + \frac{1}{2} \sum_{j=1}^{m-1} \sum_{i=1}^{d} \left( \frac{\Delta X_j^i - \widehat{\Delta X}_j^i}{\sigma_i} \right)^2, \tag{23}$$

where $\mathcal{I}_n(X)^{(i_1 \cdots i_n)} = \int_{t_n=0}^{1} \cdots \int_{t_1=0}^{t_2} dX_{t_1}^{i_1} \cdots dX_{t_n}^{i_n}$, $\Delta X_j^i = X_{j+1}^i - X_j^i$, $\widehat{\Delta X}_j^i$ are the first-guesses for them, and $\sigma_i^2 > 0$ are the variances for the first-guess. The last term in the right hand side represents the prior knowledge, where the first-guess in this case is path segments sampled from the straight path from $(0,0,0)$ to $(\mu^{(1)}, \mu^{(2)}, \mu^{(3)})$ at equal intervals.

Under the above setup, we minimize the cost function with respect to $X$. By utilizing the adjoint operation for the signature transform, $(\partial I_k / \partial X_j)^\top$, we can solve the minimization problem with a gradient method.